\documentclass[prl,aps,twocolumn,
               showpacs,preprintnumbers,
               floatfix,numbers,sort&compress,
               colorlinks,
               citecolor=blue,
               linkcolor=blue]{revtex4}
\usepackage{amsmath,amssymb,multirow,graphicx,bm,psfrag}
\newcommand{\D}[1]{#1\textsc{d}}
\newcommand{\Ref}{Ref.\ }
\newcommand{\Refs}{Refs.\ }
\newcommand{\textos}[1]{#1}

\usepackage{mciteplus}
\mciteErrorOnUnknownfalse
\newcommand{\joinref}{,*}

\newcommand{\exclude}[1]{}
\newcommand{\MIT}{\textsc{mit}}
\newcommand{\YITP}{\textsc{yitp}}
\newcommand{\LOFF}{\textsc{loff}}
\newcommand{\FFLO}{\textsc{fflo}}
\newcommand{\SF}{\textsc{sf}}
\newcommand{\DFT}{\textsc{dft}}

\newcommand{\SLDA}{\textsc{slda}}
\newcommand{\ASLDA}{\textsc{aslda}}

\newcommand{\BCS}{\textsc{bcs}}
\newcommand{\QCD}{\textsc{qcd}}
\newcommand{\DVR}{\textsc{dvr}}
\newcommand{\LO}{\textsc{lo}}
\newcommand{\PP}{\textsc{pp}}
\newcommand{\FF}{\textsc{ff}}
\newcommand{\BdG}{BdG}

\newcommand{\sNSF}{{\scriptscriptstyle N-SF}}
\newcommand{\sLON}{{\scriptscriptstyle LO-N}}
\newcommand{\sLOSF}{{\scriptscriptstyle LO-SF}}

\newcommand{\vect}[1]{{\bm #1}}
\providecommand{\d}{}
\renewcommand{\d}{\mathrm{d}}
\DeclareMathOperator{\sign}{sign}

\newcommand*\savefootnote[2]{%
  \newsavebox{#1}%
  \sbox{#1}{\protect\footnote{#2}}}

\begin{document}
\title{A Unitary Fermi Supersolid: The Larkin-Ovchinnikov Phase}
 
\author{Aurel Bulgac}
\author{Michael McNeil Forbes}
\email[E-mail:~]{mforbes@alum.mit.edu}
\affiliation{Department of Physics, University of Washington, Seattle,
  WA 98195--1560, USA}

\date{18 November 2008}
\preprint{NT@UW-08-09}

\pacs{03.75.Ss, 03.75.Hh, 05.30.Fk, }

\keywords{Asymmetric Fermi gas, unitary Fermi gas, superfluidity,
  thermodynamics, phase structure, cold atoms, supersolid,
  LOFF,FFLO,LO,FF,Larkin-Ovchinnikov,Fulde-Ferrell}

\begin{abstract}
  We present strong theoretical evidence that a Larkin-Ovchinnikov
  (\LOFF/\FFLO) pairing phase is favoured over the homogeneous
  superfluid and normal phases in three-dimensional unitary Fermi
  systems.  Using a Density Functional Theory (\DFT) based on the
  latest quantum Monte-Carlo calculations and experimental results, we
  show that this phase is competitive over a large region of the phase
  diagram.  The oscillations in the number densities and pairing field
  have a substantial amplitude, and a period some 3 to 10 times the
  average interparticle separation.  Within the \DFT, the transition
  to a normal polarized Fermi liquid at large polarizations is smooth,
  while the transition to a fully-paired superfluid is abrupt.
\end{abstract}

\maketitle
The \BCS\ mechanism for fermionic superfluidity is rooted in the
notion of pairing: Can superfluidity survive in polarized systems with
unequal numbers?  This question remains largely unanswered, even
though it is fundamental to many forms of matter, including
superconductors, nuclear matter, and high density \QCD.  The prospect
of observing exotic polarized superfluids has been revived in
two-component cold-atoms gases with \textit{s}-wave interactions,
especially in the unitary limit where the scattering length diverges:
$|a|\rightarrow \infty$.  Here, the physics is universal, and
inherently strongly coupled, depending solely on the densities.  These
systems are experimentally tenable (see \Ref \cite{KZ:2008} for a
review), and exhibit a remarkable diversity of polarized phases.

Clogston and Chandrasekhar~\cite{Clogston:1962\joinref
  Chandrasekhar:1962} noted that the normal phase competes with \BCS\
superfluidity when the chemical potential difference between the
species becomes comparable to the energy gap.  Kohn and
Luttinger~\cite{KL:1965}, however, showed that interactions render
Fermi surfaces unstable at sufficiently low temperatures,
suggesting pairing of higher angular momenta.  This effect is
exponentially suppressed in weak coupling, but may be strong enough in
unitary gases to support symbiotic $p$-wave
superfluids~\cite{Bulgac:2006gh\joinref BY:2008}.  Another proposal by
Fulde and Ferrell (\FF) \cite{FF:1964}, and Larkin and Ovchinnikov
(\LO) \cite{LO:1965} -- anisotropic/inhomogeneous polarized
superfluids, widely referred to as \LOFF\ or \FFLO\ states -- have
also been vigorously sought (see~\cite{Combescot:2008\joinref
  Casalbuoni:2003wh} for reviews), but firm results have been sparse:
Experimentally there have been claims of quasi--two-dimensional \FFLO\
states in heavy-fermion
superconductors~\cite{kakuyanagi:047602\joinref BMOGSTPS:2002\joinref
  Radovan:2003lr\joinref MNPMSTSS:2006\joinref NMSTSS:2007\joinref
  YUCTSVG:2007\joinref
  Yonezawa;Kusaba;Maeno;Auban-Senzier;Pasquier...:2008-04}, but no
\D{3} realizations have been reported.  Other proposals include
deformed Fermi surfaces~\cite{Muther:2002mc} and gapless (breached
pair) superfluids~\cite{Liu:2002gi\joinref Gubankova:2003uj\joinref
  Forbes:2004cr}.

We present here strong evidence that an inhomogeneous
Larkin-Ovchinnikov (\LO) state~\cite{LO:1965} may be realized in cold
polarized unitary Fermi gases.  Our approach is novel in several
respects: 1) it is the first calculation to find a completely
self-consistent \LO\ solution in three dimensions; 2) the calculation
is based on a Density Functional Theory (\DFT) incorporating the best
Monte-Carlo calculations and measurements of the unitary Fermi gas;
and 3) includes both pairing and self-energy correlations.

Previous calculations of \LO\ states have not been fully
self-consistent, often relying on approximate forms of spatial
variations, or uncontrolled Ginzburg-Landau expansions (see e.g.~\Refs
\cite{Combescot:2008\joinref
  Casalbuoni:2003wh,YY:2007,Matsuo;Higashitani;Nagato;Nagai:1998-01}).
Furthermore, self-consistent treatments are typically based on
mean-field or Bogoliubov--de-Gennes (\BdG) calculations, which do not
properly account for many-body effects such as the
Gorkov--Melik-Barkhudarov corrections~\cite{GM-B:1961\joinref
  HPSV:2000} that lead to significant decreases in the pairing gap.
Finally, most calculations account for only the pairing condensation
energy, which is exponentially suppressed in weak-coupling, while the
\LO\ state has density variations that can significantly change the
unsuppressed normal correlation energy (``Hartree'' terms).
Mean-field and \BdG\ calculations neglect these crucial correlation
contributions: Without them, \LO\ states are not competitive at
unitarity.

According to the theorems of Hohenberg and Kohn, a \DFT\ exists for
any system of fermions.  At unitarity, the structure of the functional
is strongly constrained by dimensional arguments, and thus its
determination is greatly simplified.  The remarkable accuracy of this
approach for symmetric systems -- as demonstrated in \Ref
\cite{Bulgac:2007a} -- gives us the confidence to extend the approach
to polarized systems.  To model the polarized unitary Fermi gas, we
use an asymmetric (\ASLDA) generalization of the superfluid local
density approximation (\SLDA) employed in \Ref \cite{Bulgac:2007a},
expressed in terms of the following densities
\begin{gather}
  \label{eq:Densities}
  \begin{aligned}
    n_{a}(\vect{r}) &= \sum_{E_{n}<0}|u_{n}(\vect{r})|^2, &
    n_{b}(\vect{r}) &= \sum_{0<E_{n}}|v_{n}(\vect{r})|^2,\\
    \tau_{a}(\vect{r}) &= \sum_{E_{n}<0}|\vect{\nabla} u_{n}(\vect{r})|^2, &
    \tau_{b}(\vect{r}) &= \sum_{0<E_{n}}|\vect{\nabla} v_{n}(\vect{r})|^2,
  \end{aligned}\nonumber\\
  \nu (\vect{r}) = \frac{1}{2}
  \sum_{E_{n}}\sign(E_n)u_{n}(\vect{r})v_{n}^{*}(\vect{r}),
\end{gather}
where $u_{n}(\vect{r})$, $v_{n}(\vect{r})$, and $E_{n}$ are the
quasiparticle wave-functions and energies \cite{Bulgac:2007a}. We use
the same functional form as \Ref \cite{Bulgac:2007a}, but allow the
parameters to depend on the local asymmetry
$x(\vect{r})=n_{b}(\vect{r})/n_{a}(\vect{r})$.  The resulting \ASLDA\
energy density $\mathcal{E}(\vect{r})$ has the form
\begin{multline}
  \label{eq:DF_ASLDA}
  \mathcal{E}(\vect{r}) =
  \frac{\hbar^2}{2m}\Bigl[
  \alpha_{a}(\vect{r})\tau_{a}(\vect{r})+ 
  \alpha_{b}(\vect{r})\tau_{b}(\vect{r}) 
  \Bigr] + g_{\text{eff}}(\vect{r})|\nu(\vect{r})|^2 +\\
  + \frac{3(3\pi^2)^{2/3}\hbar^2 }{10m}
  \Bigl[n_{a}(\vect{r})+n_{b}(\vect{r})\Bigr]^{5/3}
  \beta\Bigl[x(\vect{r})\Bigr],
\end{multline}
where $\alpha_{a}(\vect{r}) =\alpha\bigl[x(\vect{r})\bigr]$ and
$\alpha_{b}(\vect{r})=\alpha\bigl[1/x(\vect{r})\bigr]$ are the inverse
effective masses in units of $m^{-1}$ defined in terms of the
single function $\alpha(x)$, $\beta(x) = \beta(1/x)$ parametrizes the
normal interaction, and $(n_{a}+n_{b})^{1/3}/\gamma = 1/g_{\text{eff}} + \Lambda$
defines the effective coupling $g_{\text{eff}}$ that is regulated with the cutoff
$\Lambda$ as described in \Refs \cite{Bulgac:2007a,Bulgac;Forbes:2008-08}. The
forms of $\alpha(x)$ and $\beta(x)$ are well constrained by
Monte-Carlo data, as described below and in \Ref
\cite{Bulgac;Forbes:2008-08}. The equations for the quasiparticle wave-functions
$u_{n}(\vect{r})$ and $v_{n}(\vect{r})$ follow by minimizing the
grand-canonical functional
\begin{equation*}
 \Omega = -\!\!\int\!\d^{3}\vect{r}\;\mathcal{P}(\vect{r})
 = \int\!\d^{3}\vect{r}\Bigl[ 
   \mathcal{E}(\vect{r}) - \mu_a
   n_{a}(\vect{r})-\mu_b n_{b}(\vect{r})
 \Bigr], 
\end{equation*}
where $\mu_{a,b}$ are the chemical potentials corresponding to the two
fermion species, and $\mathcal{P}(\vect{r})$ is the local pressure.

\begin{figure}[tbp]
  \begin{center}
    \psfrag{x=n_b/n_a}{$x=n_{b}/n_{a}$}
    \psfrag{g(x)}{$g(x)$}
    \psfrag{0.9}{\footnotesize 0.9}
    \psfrag{1.0}{\footnotesize 1.0}
    \psfrag{1.1}{\footnotesize 1.1}
    \psfrag{0.0}{\footnotesize 0.0}
    \psfrag{0.2}{\footnotesize 0.2}
    \psfrag{0.4}{\footnotesize 0.4}
    \psfrag{0.6}{\footnotesize 0.6}
    \psfrag{0.8}{\footnotesize 0.8}
    \includegraphics[width=\columnwidth]{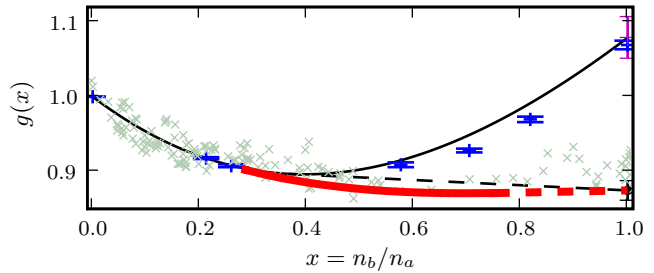}
    \caption{\label{fig:g_x} (Color online) The dimensionless convex
      function $g(x)$ \cite{Bulgac:2006cv} that defines the energy
      density $\mathcal{E}(n_a,n_b) =
      \tfrac{3}{5}\tfrac{\hbar^2}{2m}(6\pi^2)^{2/3}\left[n_{a}
        g(x)\right]^{5/3}$. The points with error-bars (blue online)
      are the Monte-Carlo data from \Refs \cite{LRGS:2006\joinref
        CRLC:2007\joinref PG:2008}.  The fully-paired solution
      $g(1)=(2\xi)^{3/5}$ is indicated to the bottom right, and the
      recent \MIT\ data~\cite{shin-2008} is shown (light $\times$) for
      comparison.  The phase separation discussed in
      Refs.~\cite{LRGS:2006\joinref CRLC:2007\joinref PG:2008} is
      shown by the Maxwell construction (thin black dashed line) of
      the $1^\text{st}$-order transition $y=y_{\sNSF}$ in
      Fig.~\ref{fig:h_y}.  The \LO\ state (thick red curve) has
      \emph{lower energy} than all pure states and phase separations
      previously discussed.  The Maxwell construction of the weakly
      $1^\text{st}$-order transition $y=y_{\sLOSF}$ in
      Fig.~\ref{fig:h_y} is shown by the thick dashed line (red).}
  \end{center}
\end{figure}

We must now specify the forms of $\alpha(x)$ and $\beta(x)$.  First we
analyze the symmetric superfluid phase as described in \Refs
\cite{Bulgac:2007a,Bulgac;Forbes:2008-08}.  By matching the
Monte-Carlo
values~\cite{CCPS:2003,Carlson:2005kg,BDMW:2008\joinref
  BDM:2008,Carlson:private_comm2} for the parameters $\xi =
\mathcal{E}_{SF}/\mathcal{E}_{FG} = 0.40(1)$, $\eta =
\Delta/\varepsilon_{F} = 0.504(24)$, and the single quasiparticle
excitation spectrum, we determine the inverse effective mass
$\alpha(1) \approx 1.09(2)$, the constant $\gamma^{-1} = -0.091(8)$,
and the implied energy of the symmetric normal phase $\xi_{N} =
\mathcal{E}_{N}/\mathcal{E}_{FG} = \alpha(1)+\beta(1) = 0.57(2)$. In
\Ref \cite{Bulgac:2007a} it was erroneously stated that $\alpha(1)$
could be extracted from the values of $\xi$ and $\eta$ alone.  A more
careful analysis shows that the quasiparticle dispersion
\cite{Carlson:2005kg} must also be fit, resulting in the modified
\SLDA\ parameters presented here.  The inverse masses $\alpha_{a,b}$
are also known for the fully-polarized gas. The majority species is
unaffected, whereas the minority species has the effective ``polaron''
mass $m/m^* \approx 1.04(3)$~\cite{LRGS:2006\joinref CRLC:2007\joinref
  PG:2008,PS:2008}, constraining the endpoints $\alpha(0) = 1$ and
$\alpha(\infty)= 0.96(3)$.  To determine the function $\beta(x)$, we
consider the energy of the interacting normal state, setting
$\Delta\equiv 0$.  This has been well constrained by Monte-Carlo
calculations~\cite{CCPS:2003,LRGS:2006\joinref CRLC:2007\joinref
  PG:2008}, and, along with the parametrization of $\alpha(x)$, a fit
to this data uniquely specifies the function $\beta(x)$ (see
Fig.~\ref{fig:g_x}).
 
In this paper, we consider only the simplest \LO\ states, with spatial
modulations in a single direction ($z$).  Unlike the \FF\
state~\cite{FF:1964}, the \LO\ state~\cite{LO:1965} does not break
time-reversal invariance, and thus belongs to a different symmetry
class, as was emphasized by Yoshida and Yip \cite{YY:2007}.  We do not
consider \FF\ states here as they are typically not competitive with
the \LO\ states.  (\FF\ states break time-reversal invariance, and
require additional terms in~(\ref{eq:DF_ASLDA}) to restore Galilean
invariance~\cite{Bulgac:2007a,Bulgac;Forbes:2008-08}.)

The self-consistency equations are solved by discretizing the
Hamiltonian along $z$ with a discrete variable representation (\DVR)
basis~\cite{LCCMP:2002} of period $L$, while integrating over the
perpendicular momenta and the Bloch states. Our quasiparticle
wave-functions thus satisfy the conditions $u_{n}({x,y,z+L}) =
e^{i\phi_{n}}u_{n}(x,y,z)$, $v_{n}({x,y,z+L}) =
e^{i\phi_{n}}v_{n}(x,y,z)$, and are plane waves in the $(x,y)$ plane.
We minimize the truncation error due to the finite \DVR\ basis set by
using a smoothed version of the hybrid
strategy~\cite{PhysRevA.59.3816}, summing discrete states with
energies less than a cutoff $E_{c}$, while integrating over the remaining
higher-energy semi-classical states.

We start by specifying chemical potentials $\mu_{a}$ and $\mu_{b}$,
and an ansatz for $\Delta(z) \propto \sin(2\pi z/L)$ containing a
node at $z=0$, and then use a Broyden iteration scheme
\cite{Baran;Bulgac;Forbes;Hagen;Nazarewicz...:2008-05} to
find a self-consistent solution.  The choice of basis and iteration
preserve the node at $z=0$, converging to either a self-consistent
\LO\ state, or degenerating to a homogeneous normal state with
$\Delta(z) = 0$ everywhere.

The resulting self-consistent states depend on the external parameter
$L$.  To find the physical \LO\ state, we vary $L$ to find the
spontaneously chosen length scale $L=L_{LO}$ that minimizes the
potential $\Omega$ (maximizes the average pressure $\mathcal{P}$).
The search is aided by the relationship
$L\,\partial\mathcal{P}/\partial L = 2\mathcal{E}-3\mathcal{P}$
between $L$, the average pressure, and the energy
density~\cite{Bulgac;Forbes:2008-08}, ensuring the unitary relationship
$\mathcal{P} = \tfrac{2}{3}\mathcal{E}$ is satisfied by the physical
state.

\savefootnote{\mynote}{Our units are chosen so that
  $\mu_{-}$ is held fixed as is for trapped systems.  Thus,
  $\mu_{+} = \mu_{-}(1+y)/(1-y)$.  We normalize everything in
  terms of the density $n_{0} = n_{a}|_{y=y_{\sLOSF}}$,
  interparticle spacing $l_{0} = n^{-1/3}|_{y=y_{\sLOSF}}$,
  and pairing gap $\Delta_{0}$ of the \SF\ phase at the
  $1^{\text{st}}$-order transition $y=y_{\sLOSF}$ where the
  \LO\ phase ends.}
\begin{figure}[tbp]
  \begin{center}
    \psfrag{y=mu_b/mu_a}{$y=\mu_{b}/\mu_{a}$}
    \psfrag{h(y)}{$h(y)$}
    \psfrag{y_0}{$y_{0}$}
    \psfrag{y_1}{$y_{1}$}
    \psfrag{yLO}{$\!\!\!\!\!y_{\sNSF}$}
    \psfrag{yLON}{$y_{\sLON}$}
    \psfrag{yLOSF}{$y_{\sLOSF}$}
    \psfrag{Delta__}{$\Delta/\Delta_{0}$}
    \psfrag{L/l0__}{$L/l_{0}$}
    \psfrag{0}{\footnotesize 0}
    \psfrag{1}{\footnotesize 1}
    \psfrag{5}{\footnotesize 5}
    \psfrag{10}{\footnotesize 10}
    \psfrag{1.0}{\footnotesize 1.0}
    \psfrag{1.1}{\footnotesize 1.1}
    \psfrag{1.2}{\footnotesize 1.2}
    \psfrag{1.3}{\footnotesize 1.3}
    \psfrag{0.0}{\footnotesize 0.0}
    \psfrag{0.1}{\footnotesize 0.1}
    \psfrag{-0.2}{\footnotesize -0.2}
    \psfrag{-0.4}{\footnotesize -0.4}
    \psfrag{-0.6}{\footnotesize -0.6}
    \includegraphics[width=\columnwidth]{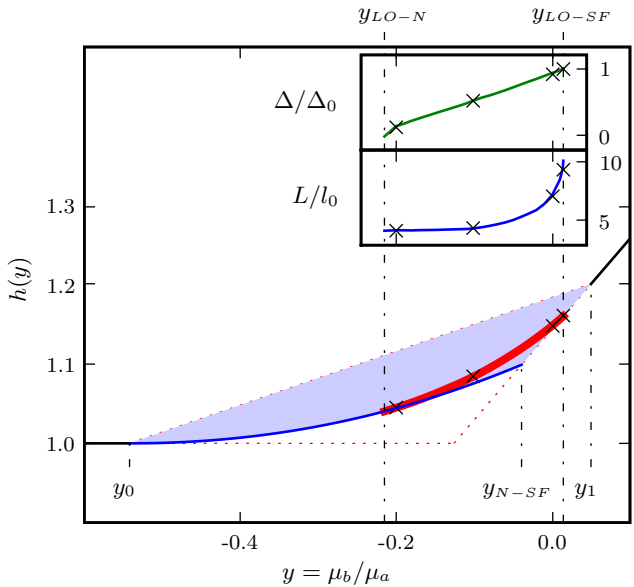}
    \caption{\label{fig:h_y} (Color online) The dimensionless convex
      function $h(y)$ \cite{Bulgac:2006cv} that defines the average
      pressure $\mathcal{P}(\mu_a,\mu_b)=
      \tfrac{2}{5}\smash{\left(\tfrac{2m}{\hbar^2}\right)^{3/2} 
        \left[\mu_a h(y)\right]^{5/2}}\!/(6\pi^2)$ is constrained to
      the thin dotted triangular region~\cite{Bulgac:2006cv}.  The
      interacting normal state pressure~\cite{LRGS:2006\joinref
        CRLC:2007\joinref PG:2008} defining our \ASLDA\ functional
      constrains this further (thin blue line), and displays a
      $1^\text{st}$-order transition at $y_{\sNSF}$ where normal and
      \SF\ phases could coexists (Maxwell construction in
      Fig.~\ref{fig:g_x}).  The \LO\ state has an even higher pressure
      (thick red line), replacing much of this region, including the
      former $y_{\sNSF}$ transition.  The $y$ dependence of the
      amplitude of the pairing field $\Delta = \max\{|\Delta(z)|\}$
      and the period $L$ are shown inset.  Sample profiles for the
      states marked $\times$ are shown in Fig.~\ref{fig:delta_x}.
      Units are fixed in terms of $\mu_{-} = (\mu_{a}-\mu_{b})/2$
      \usebox{\mynote}.}
  \end{center}
\end{figure}
\begin{figure}[t]
  \begin{center}
    \psfrag{Delta(z)}{$\Delta(z)/\Delta_{0}$}
    \psfrag{n(z)/n0}{$n(z)/n_{0}$}
    \psfrag{z/l0}{$z/l_{0}$}
    \psfrag{-1}{\footnotesize -1}
    \psfrag{0}{\footnotesize 0}
    \psfrag{1}{\footnotesize 1}
    \psfrag{0.0}{\footnotesize 0.0}
    \psfrag{0.5}{\footnotesize 0.5}
    \psfrag{1.0}{\footnotesize 1.0}
    \psfrag{-5}{\footnotesize -5}
    \psfrag{0}{\footnotesize 0}
    \psfrag{5}{\footnotesize 5}
    \includegraphics[width=\columnwidth]{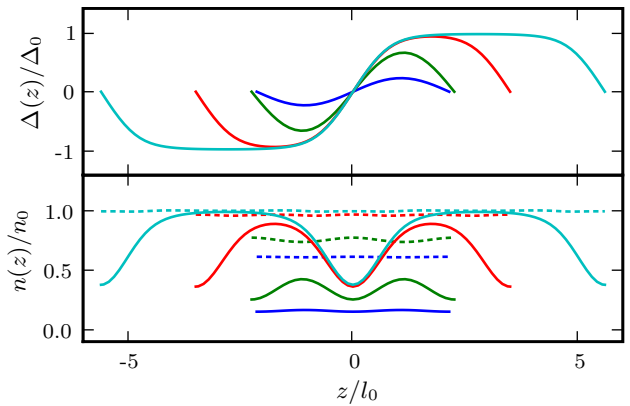}
    \caption{\label{fig:delta_x} (Color online) A single \LO\ period
      showing the spatial dependence of the pairing field $\Delta(z)$
      (top) and the number densities of the majority (dotted) and
      minority (solid) species (bottom) at the values of $y \in
      (y_{LO-N},y_{LO-SF})$ marked by $\times$ in Fig.~\ref{fig:h_y}.
      Units are fixed in terms of $\mu_{-}$ \usebox{\mynote}.}
  \end{center}
\end{figure}
At unitarity, one may fully characterize all stable phases by the
single parameter $y=\mu_{b}/\mu_{a}$ as described in \Ref
\cite{Bulgac:2006cv}.  We use the grand canonical ensemble, where only
pure phases appear, and which properly accounts for the phase
separation that occurs at 1$^{\text{st}}$-order transitions (kinks in
$h(y)$, discontinuities in $h'(y)$: see Fig.~\ref{fig:h_y}).  We start
by describing the homogeneous and isotropic states supported in the
\ASLDA\ functional: For $y < y_{0}$ \cite{Bulgac:2006cv}, the system
is a fully-polarized non-interacting Fermi gas ($N_{a}$); between
$y_{0} < y < y_{\sNSF}$ the highest pressure corresponds to a
partially polarized two-component Fermi gas; and above
$y_{\sNSF}<y<1$, the fully-paired superfluid (\SF) has the highest
pressure.  The point $y_{\sNSF}$, where the pressures of the partially
polarized normal and fully-paired superfluid states are equal, is
where the phase-separation discussed in \Refs \cite{LRGS:2006\joinref
  CRLC:2007\joinref PG:2008} would occur.  Here, the competition to
\LO\ from the normal and superfluid states is minimized, and the \LO\
state is most likely to occur.  For $y>1$, the picture is reversed
with the species $a\leftrightarrow b$ exchanged.  Our \ASLDA\
parametrization does not admit any stable homogeneous gapless
superfluid (breached pair) states~\cite{Liu:2002gi\joinref
  Gubankova:2003uj\joinref Forbes:2004cr}.

As shown in Fig.~\ref{fig:h_y}, we find competitive \LO\ solutions for
a large range of the parameter $y \in (y_{\sLON},y_{\sLOSF})$ with
finite periods in the range $L_{LO-N} \leq L \leq L_{LO-SF}$.  At
$y_{\sLON}$, the transition appears to be second order, with
$\max\{|\Delta(z)|\} \rightarrow 0$ vanishing smoothly from the \LO\
phase to the normal phase, while at $y_{\sLOSF}$, the order parameter
abruptly looses its spatial oscillations at a finite period
$L_{LO-SF}$.  Due to the presence of a node in $\Delta(z)$, the only
possibility for a smooth transition here would be for the period to
diverge $L_{\LO-\SF}\rightarrow \infty$, thus this transition appears
to be weakly first order.  The remaining normal states between $y_{0}
< y < y_{\sLON}$ would be susceptible to the Kohn-Luttinger
instability, and are candidates for the symbiotic \textit{p}-wave
superfluids discussed in \Refs \cite{Bulgac:2006gh\joinref BY:2008}.
To study this possibility requires an extension of the \ASLDA.

Fig.~\ref{fig:delta_x} shows the typical structure of a \LO\ state.
The pairing amplitude increases smoothly from zero at $y_{\sLON}$,
where the profile is almost sinusoidal, to a critical value slightly
less than $\Delta_{0}$ at $y_{\sLOSF}$, while the minority component
exhibits large oscillations that break translation invariance, giving
the \LO\ state the crystalline properties of a quantum solid.  The
majority component exhibits much smaller oscillations because the
larger local kinetic energy density suppresses gradients.  These
fluctuations induce large oscillation in the mean-field potentials
(not shown), and have a significant impact on the normal correlation
energy.  For this reason, all the terms in the energy density
functional are critical for a proper description of the \LO\ phase.

If no other phases compete, one should observe that the \LO-\SF\
transition coincides with the termination point $y_{1}$ of the
partially polarized phases \PP$_{a}$~\cite{Bulgac:2006cv}: $y_{\sLOSF}
= y_{1}$.  A more complicated crystalline \LO\ state with modulations
in all three directions may further increase the average pressure,
making $y_{1} > y_{\sLOSF}$.  Current errors of the current
Monte-Carlo calculations and experiments do not allow us to
distinguish between these two cases.  These results are summarized in
Fig.~\ref{fig:h_y}, where it can be seen that these periodic \LO\
solutions occupy a substantial portion of the phase diagram, and lead
to a significant increase in the average pressure.  The Legendre
transformed results have been included in Fig.~\ref{fig:g_x} to
facilitate comparisons with the Monte-Carlo
data~\cite{LRGS:2006\joinref CRLC:2007\joinref PG:2008} and
recent experiments~\cite{shin-2008}.

In conclusion, we have shown that the \ASLDA\ provides a valuable tool
for quantitatively evaluating inhomogeneous phases.  By incorporating
the latest non-perturbative data about unitary Fermi gases, we have
presented strong evidence that a new form of matter, such as a
crystalline \LO\ phase, is waiting to be found in the partially
polarized regime of cold unitary Fermi gases. This would be the first
example of a Fermi supersolid at unitarity, and with the large pairing
gap, there is a good chance of successfully studying this state with
cold atoms.  In the experiments to date, the shells where \LO\ phases
may exist are too thin to allow for a complete \LO\ period $L$.
However, traps can be adjusted so that the \LO\ phase will occupy a
larger spatial region, allowing for several \LOFF\ oscillations to
occur.  Unlike \LO\ in weak coupling, the amplitude of the density
fluctuations in the minority component is large and comparable to
that of vortices~\cite{BY:2003\joinref ZA-SSSK:2005lr} in unitary
gases.  This will provide the most direct signature of unitary
supersolid matter, and an clean way to study the \LOFF\ phase.

\acknowledgments M.~M.~F. would like to thank R.~Sharma for help with
the code, and acknowledges the \YITP\ workshop ``New Frontiers in
\QCD\ \textos{2008}''.  The authors would also like to thank
S.~Giorgini and Y.~Shin for providing the data in Fig.~\ref{fig:g_x},
and the US Department of Energy for support under Grants
No.\ DE-FG02-97ER41014 and No.\ DE-FC02-07ER41457.
\ifx\mcitethebibliography\mciteundefinedmacro
\PackageError{apsrevM.bst}{mciteplus.sty has not been loaded}
{This bibstyle requires the use of the mciteplus package.}\fi

\end{document}